\documentclass[a4paper,twocolumn,11pt,unpublished]{quantumarticle}

\pdfoutput=1

\makeatother

\usepackage{amsmath} 
\usepackage{amsfonts}
\usepackage{graphicx}
\usepackage{enumitem}
\usepackage{listings}
\usepackage{color}
\usepackage{subcaption}

\usepackage[
  style=numeric-comp,
  sorting=none,
  defernumbers=false,
  sortcites=false,
  backend=biber,
  doi=false,
  url=false,
  hyperref=auto,
  giveninits=true,
  uniquename=false,
  maxbibnames=6,
  minbibnames=6
]{biblatex}
\usepackage{hyperref}

\DeclareFieldFormat[article]{title}{#1}
\DeclareFieldFormat[inproceedings]{title}{#1}
\DeclareFieldFormat[incollection]{title}{#1}
\renewbibmacro{in:}{}

\DeclareFieldFormat{pages}{#1}
\DeclareFieldFormat{pagetotal}{#1}

\renewbibmacro*{journal+issuetitle}{%
  \usebibmacro{journal}%
  \setunit*{\addspace}%
  \usebibmacro{volume+number+eid}%
  \setunit{\addcomma\space}%
  \newunit}

\renewbibmacro*{volume+number+eid}{%
  \printfield{volume}%
  \setunit*{\addnbthinspace}
  \printfield{eid}}

\renewbibmacro*{issue+date}{%
  \newunit}

\renewbibmacro*{doi+eprint+url}{%
  \iftoggle{bbx:doi}
    {\printfield{doi}}
    {}%
  \newunit\newblock
  \iftoggle{bbx:eprint}
    {\usebibmacro{eprint}}
    {}%
  \newunit\newblock
  \iftoggle{bbx:url}
    {\usebibmacro{url+urldate}}
    {}%
  \setunit{\addspace}%
  \printtext[parens]{\printfield{year}}
}

\renewbibmacro*{title}{%
  \ifboolexpr{
    test {\iffieldundef{url}}
    and
    test {\iffieldundef{doi}}
  }
  {
    \printtext[title]{%
      \printfield[titlecase]{title}%
      \setunit{\subtitlepunct}%
      \printfield[titlecase]{subtitle}%
    }%
  }
  {
    \iffieldundef{doi}
      {
        \printtext[title]{%
          \href{\thefield{url}}{%
            \printfield[titlecase]{title}%
            \setunit{\subtitlepunct}%
            \printfield[titlecase]{subtitle}%
          }%
        }%
      }
      {
        \printtext[title]{%
          \href{https://doi.org/\thefield{doi}}{%
            \printfield[titlecase]{title}%
            \setunit{\subtitlepunct}%
            \printfield[titlecase]{subtitle}%
          }%
        }%
      }%
  }%
}

\usepackage[svgnames]{xcolor}
\hypersetup{
    hidelinks,
    colorlinks=true,
    breaklinks=true,
    citecolor=MediumPurple,
    filecolor=LimeGreen,
    linkcolor=MediumBlue,
    urlcolor=MediumPurple
}

\newcommand{\ket}[1]{|#1\rangle}

\usepackage[most]{tcolorbox}

\newcommand{\mpl}{Max Planck Institute for the Science of Light, Erlangen, Germany}
\newcommand{\jena}{Institut für Festkörpertheorie und Optik, Friedrich-Schiller-Universität Jena, Jena, Germany}
\newcommand{\tuebingen}{Machine Learning in Science Cluster, Department of Computer Science, Faculty of Science, University of Tuebingen, Germany}

\begin{document}

\title{Automated Discovery of Non-local Photonic Gates}

\author{Sören Arlt}
\email{soeren.arlt@uni-tuebingen.de}
\affiliation{\tuebingen}
\affiliation{\mpl}

\author{Mario Krenn}
\email{mario.krenn@uni-tuebingen.de}
\affiliation{\tuebingen}

\author{Xuemei Gu}
\email{xuemei.gu@uni-jena.de}
\affiliation{\jena}

\maketitle

\begin{abstract}
Interactions between quantum systems enable quantum gates, the building blocks of quantum information processing. In photonics, direct photon-photon interactions are too weak to be practically useful, so effective interactions are engineered with linear optics and measurement. A central challenge is to realize such interactions non-locally, i.e., between photons that remain spatially separated.
We present experimental proposals for several essential non-local multiphoton quantum gates that act on spatially separated photons, in both qubit and high‑dimensional qudit systems. All solutions were discovered by the AI-driven discovery system called \texttt{PyTheus}. Rather than using pre-shared entanglement or Bell state measurements, our gates use as a resource quantum indistinguishability by path identity -- a technique that exploits coherent superpositions of the photon-pair origins. While analyzing these solutions, we uncovered a new mechanism that mimics much of the properties of quantum teleportation, without shared entanglement or Bell-state measurements. Technically, our results establish path indistinguishability as a practical resource for distributed quantum information processing; conceptually, they demonstrate how automated discovery systems can contribute new ideas and techniques in physics.
\end{abstract}

\section{Introduction}

Photons are widely regarded as ideal candidates for quantum information carriers due to their negligible decoherence and the relative ease of implementing single-qubit operations. Their weak interaction with the environment allows quantum states to remain coherent over long distances, both in free space and optical-fiber-based quantum networks. However, this feature comes with a downside -- it is extremely challenging to perform multi-photon gates. 

Direct photon-photon interactions in free space are basically negligible~\cite{rinderknecht2025measuring}, and their interactions in materials require exceptionally high nonlinearities~\cite{langford2011efficient} or intricate mediation via precisely controlled atoms~\cite{hacker2016photon}. Fortunately, effective photon-photon interactions can be realized through linear 
optics combined with heralding or post-selection based on specific detector click 
patterns~\cite{gasparoni2004realization, gao2020computer, bao2007optical, li2021heralded, liu2024nonlocal}. Such measurement-based approaches enable 
multi-photon gates, which are essential for linear optical quantum information 
processing~\cite{knill2001scheme}. These gates usually require photons to either meet at one location (e.g., a beam splitter) or to share entangled ancilla states or Bell-state measurements.


Recently, a different physical principle -- path identity and the indistinguishability of photon origins -- has emerged as an alternative resource for photonic quantum information processing~\cite{krenn2017entanglement, krenn2017quantum, hochrainer2022quantum}. For example, various state generation protocols have been proposed that can generate complex quantum states~\cite{gu2019quantum, gu2019quantum3}, even in non-local way \cite{arlt2025automated}. Moreover, a series of recent experiments have demonstrated the feasibility of path identity for state generation~\cite{kysela2020path, bao2023very, hu2025observation, wang2024entangling} and have observed the underlying multi-photon interference effects for the first time~\cite{qian2023multiphoton, feng2023chip, wang2025violation}.

\begin{figure*}[!t]
  \centering
  \includegraphics[width=\textwidth]{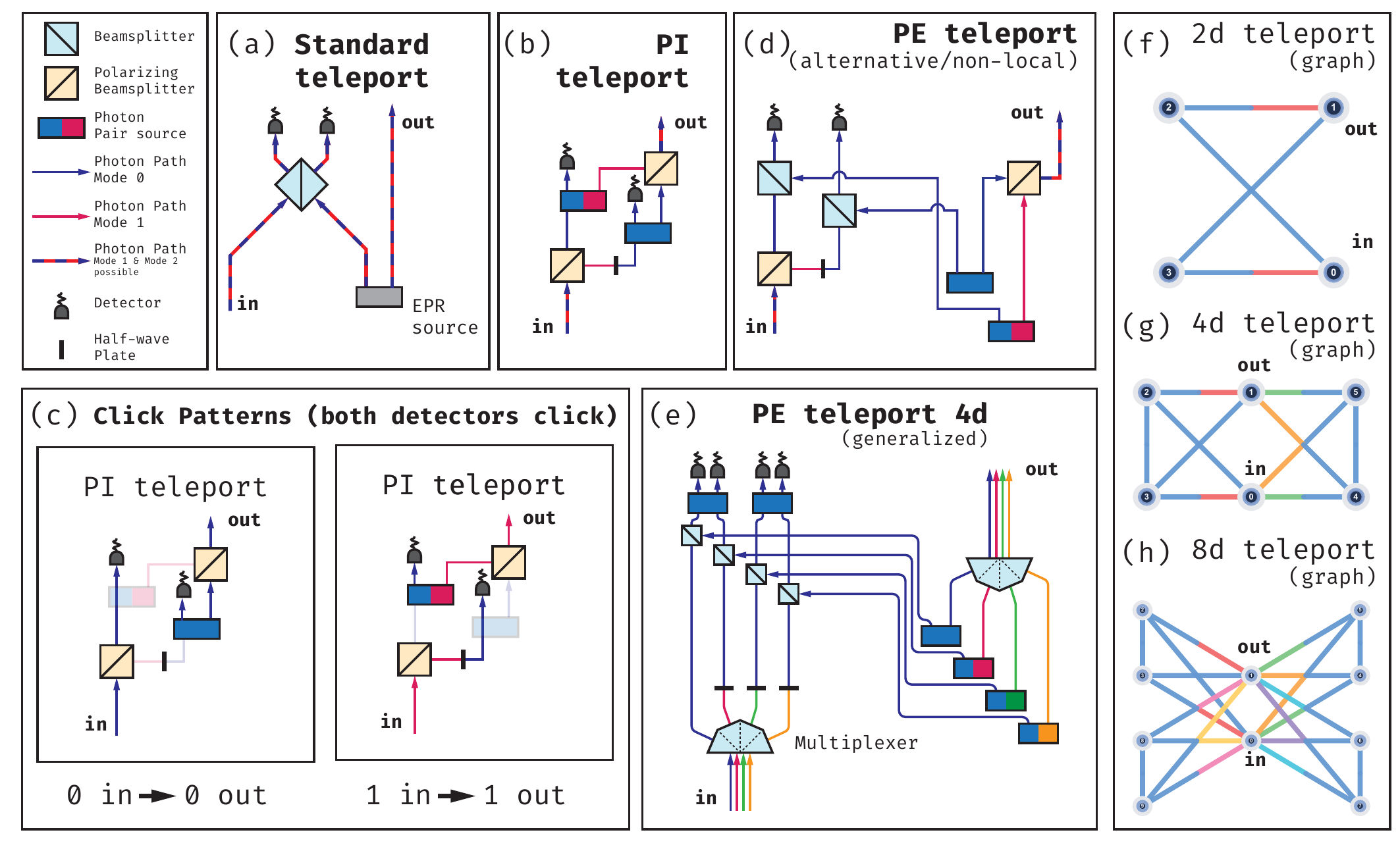}
  \caption{\textbf{Quantum teleportation} as a core component of the \texttt{SWAP} gate. (a) The standard quantum teleportation protocol, based on a shared entangled pair and a Bell-state measurement~\cite{bennett1993teleporting,bouwmeester1997experimental}. (b) A variant based on \textit{path identity} (PI)~\cite{wang2024entangling}, where the quantum information of a photon is transferred to the output, while the photon itself does not propagate to the output but is instead measured. This variation was discovered accidentally during the automated search for a \texttt{SWAP} gate. (c) The detector click patterns corresponding to the PI analogue of teleportation in (b). (d) A variation of the PI-based teleportation, enabled by removing the path information through quantum erasure (we label as PE teleportation). (e) A high-dimensional generalization of PE teleportation. (f-h) All corresponding solutions were discovered by \texttt{PyTheus}~\cite{ruiz2023digital} using the graph-based representation of quantum optics~\cite{krenn2017quantum,gu2019quantum}. These graphs represent the underlying logic of the quantum correlations encoded in the experiments.}
  \label{fig:teleport}
\end{figure*}
Here, we show that indistinguishability by path identity can also be used to implement non-local photonic gates — without the photons ever interacting, sharing a common location, or requiring the need to share entangled ancillary photons. All of the solutions are \textit{feed-forwardable}~\cite{gasparoni2004realization}, meaning that they could be used as part of quantum circuit and do not need any measurements or conditioning directly after the gate (rather, only after the circuit). This makes them practically applicable for operations among distant quantum devices in large quantum networks.

The proposals show that path identity and the indistinguishability of the origins of photons can be seen as a resource for photonic quantum information processing. The ideas demonstrated here were discovered by the automated design algorithm \texttt{PyTheus}~\cite{ruiz2023digital} -- demonstrating how artificial intelligence can help discovering new principles in physics.

\begin{figure*}[!t]
  \centering
  \includegraphics[width=\textwidth]{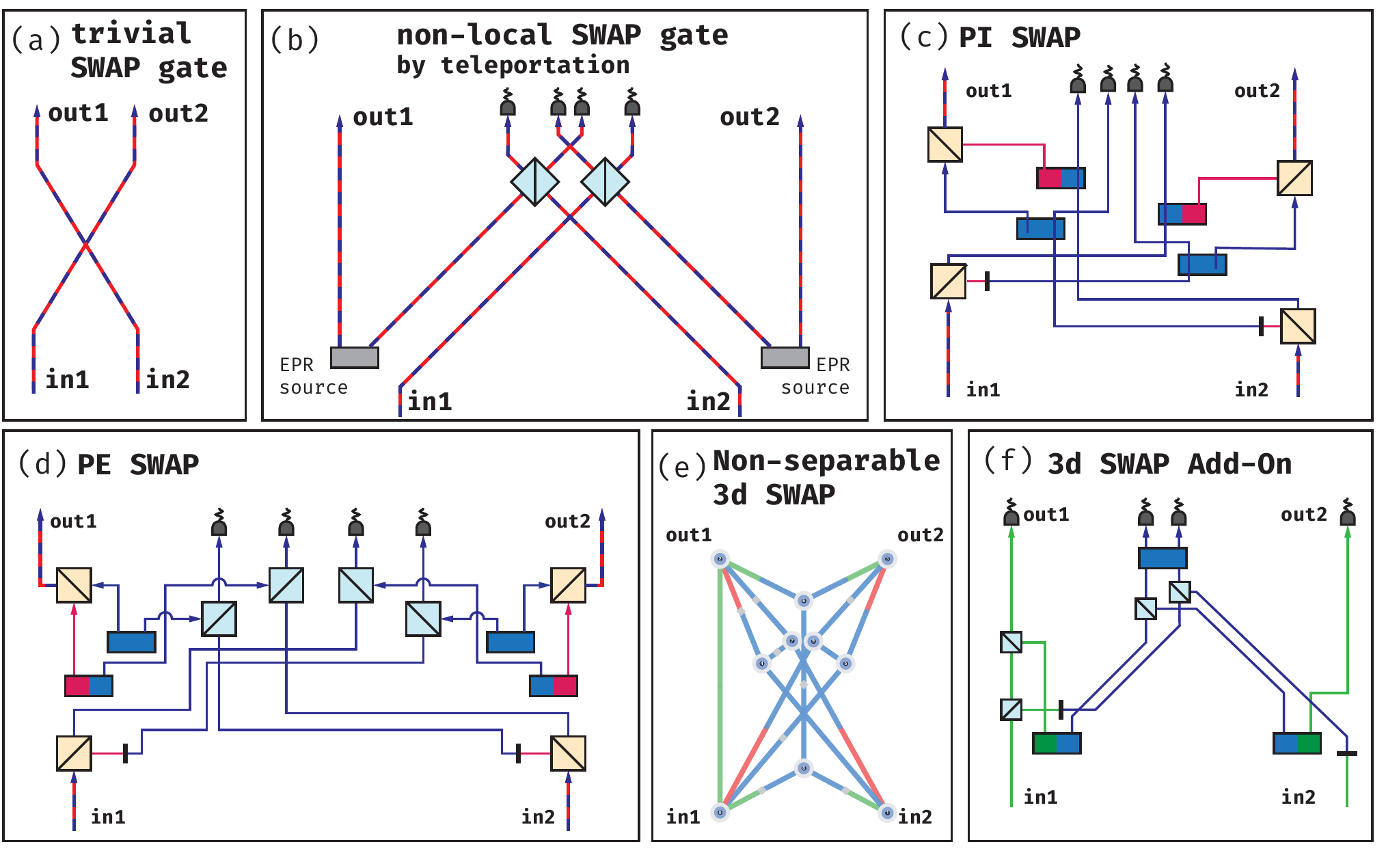}
  \caption{\textbf{Photonic \texttt{SWAP} gate}. (a) A trivial way to perform a \texttt{SWAP} operation, where both photons are physically exchanged by sending each over the whole distance. (b) A \texttt{SWAP} gate realized through two quantum teleportations~\cite{siddiqui2023swap}, without direct transmission of the photons. (c) A path-identity (PI) analogue of teleportation used to construct a \texttt{SWAP} gate. (d) An alternative scheme employing path-information quantum erasure (PE). (e-f) A three-dimensional generalization of the \texttt{SWAP} gate, shown as a graph representation and as an extension of the experiment in (d).}
  \label{fig:swap}  
\end{figure*}

\section{Representation of quantum gates via colored weighted graphs}
\texttt{PyTheus} uses an abstract graph-based representation of quantum optics that has been largely developed in~\cite{krenn2017quantum, gu2019quantum, gu2019quantum3, gu2020quantum}, and which is a native description of experiments based on path identity~\cite{zou1991induced, krenn2017entanglement, hochrainer2022quantum, bao2023very, wang2024entangling}. The core idea is to represent photon paths as graph vertices. Edges represent correlated photons (or single photons transitioning from an incoming to an outgoing path, which is useful for describing gates and can be understood through the Klyshko picture~\cite{klyshko1988simple, aspden2014experimental} or the Choi--Jamio{\l}kowski isomorphism~\cite{choi1975completely, jamiolkowski1972linear}). Each edge is assigned a color, corresponding to the photonic mode (for example, blue for $\ket{0}$, red for $\ket{1}$, and green for $\ket{2}$), and a weight, indicating the strength of the correlation. A joint detection event across all detectors is represented by a \textit{perfect matching} — a subset of edges that includes every vertex exactly once. A quantum state is then described as a coherent superposition of all such perfect matchings. For a detailed account of this formalism, we refer to~\cite{ruiz2023digital}. 

This representation is powerful because it abstracts away experimental details and focuses purely on the quantum correlation aspect of the operation, leading to readily interpretable implementations. Throughout this work, we show both graphs and concrete experimental implementations when the blueprint helps convey a critical concept. In other cases, we present only the graphs (and their perfect matchings) to describe the gate operations.

\begin{figure*}[!t]
  \centering
  \includegraphics[width=\textwidth]{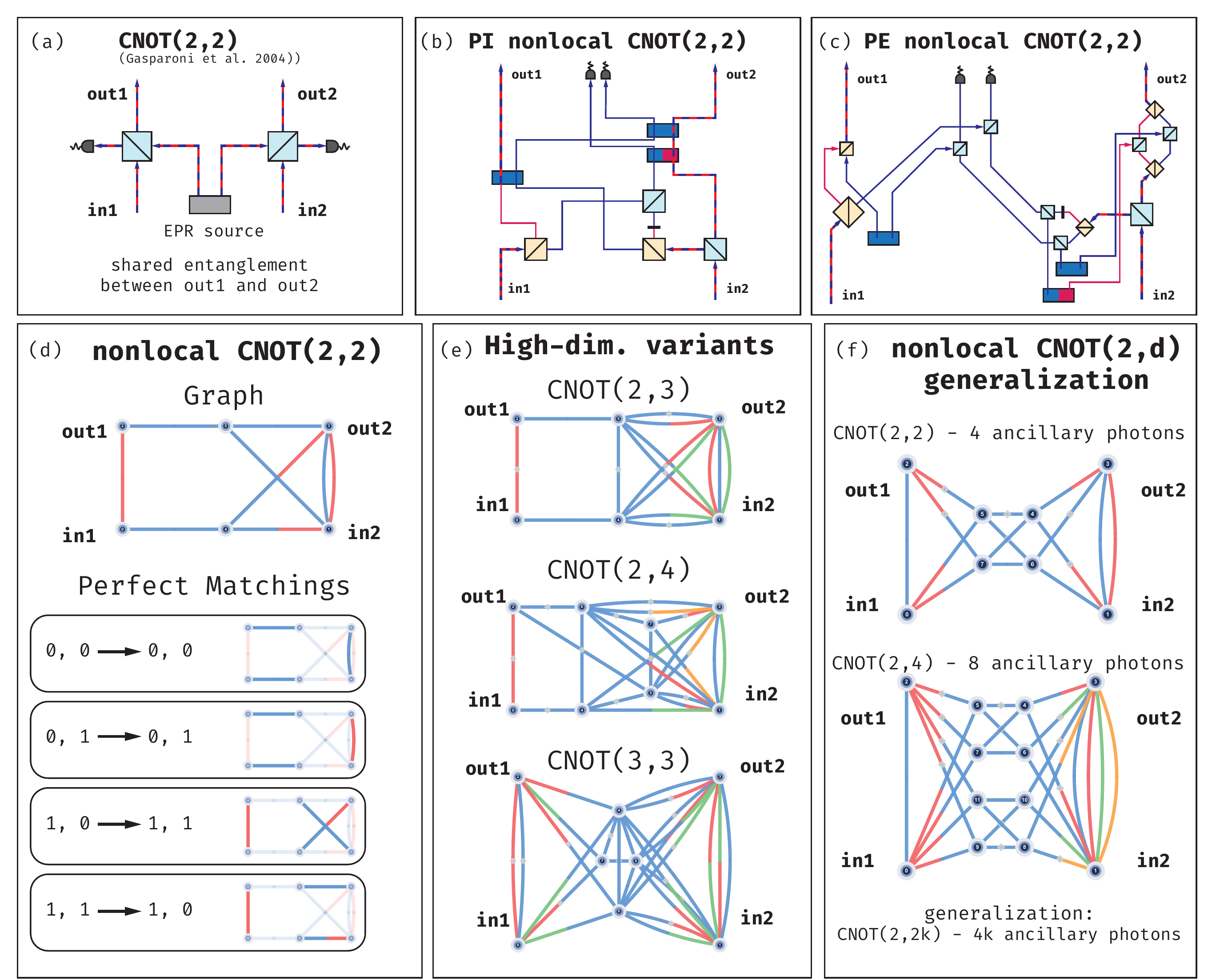}
  \caption{\textbf{Photonic \texttt{CNOT} gates}. (a) The standard implementation of a photonic \texttt{CNOT} gate, in which an entangled photon pair is shared between the two sides~\cite{gasparoni2004realization}. (b) An alternative scheme based on path identity that do not need shared entanglement between the input photons. (c) A new experimental blueprint for a \texttt{CNOT} gate realized via path-information erasure. (d) The original qubit \texttt{CNOT} solution discovered by \texttt{PyTheus}. The perfect matchings of the corresponding graph (shown below) illustrate the intrinsic working principle of the gate. (e) High-dimensional \texttt{CNOT} gates discovered by \texttt{PyTheus}. (f) Manually constructed graphs based on a generalizable pattern we identified from the machine-discovered gates. Analogous to Figs.~\ref{fig:teleport} and \ref{fig:swap}, the graphs can be directly translated into implementations based on path identity or path-information erasure.}
  \label{fig:cnot}
\end{figure*}

\section{\texttt{SWAP} gates and Teleportation}
We start by looking into SWAP gate -- an operation that exchanges the quantum information of two photons,
\begin{equation}
\widehat{\mathrm{SWAP}} \, |a, b\rangle = |b, a\rangle,
\end{equation}
where $a$ and $b$ denote qubits (or qudits). The most straightforward realization is to physically send photon $A$ to location $B$ and photon $B$ to location $A$. A conceptually more elegant approach avoids the direct transmission of the quantum carriers by using bidirectional quantum teleportation~\cite{siddiqui2023swap}, i.e., teleporting $A \to B$ and $B \to A$. 

In contrast to standard bidirectional teleportation, we discover \texttt{SWAP} gates that require neither Bell-state measurements nor pre-shared entangled photon pairs. Instead, they exploit indistinguishability by path identity as the underlying mechanism. In this process, we also discover a previously unknown and generalizable pattern for constructing \texttt{SWAP} gates in high-dimensional quantum systems~\cite{erhard2020advances}.

Interestingly, the generalizable pattern consists of two symmetric and spatially separated operations, in perfect analogy to the bidirectional quantum teleportation protocol. By this analogy, we have incidentally discovered a path-identity analogue of high-dimensional quantum teleportation, for which, so far, only solutions based on generalized Bell-state measurements were known~\cite{luo2019quantum, hu2020experimental, hu2023progress}. We refer to this new way as \textit{path-identity teleportation} (\textit{PI teleportation}, Fig.~\ref{fig:teleport} (b)).

\begin{figure*}[!t]
  \centering
  \includegraphics[width=\textwidth]{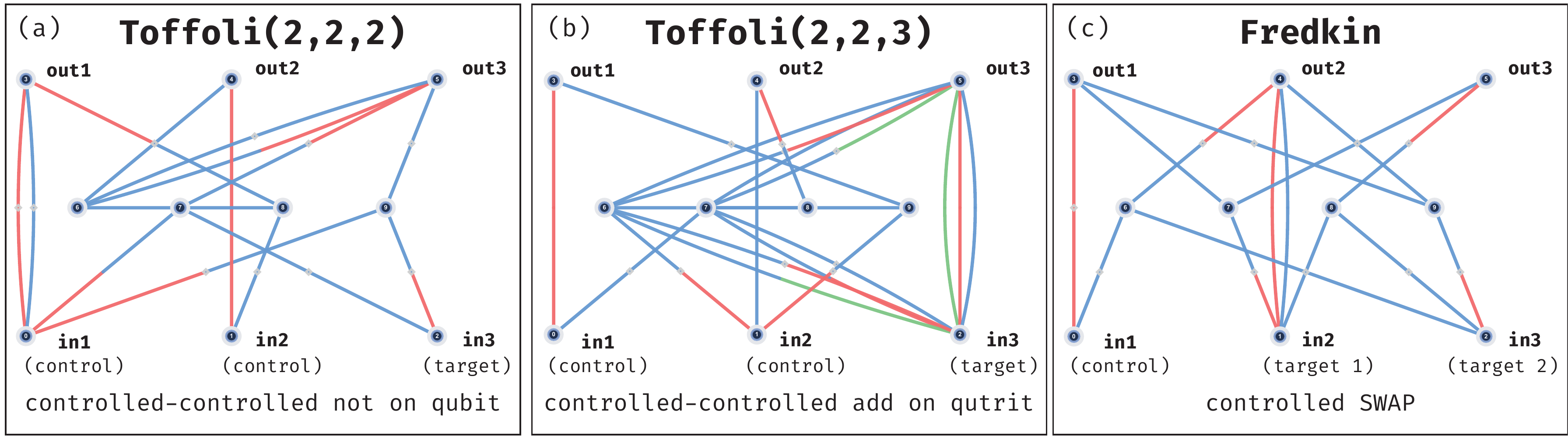} 
  \caption{\textbf{Experimental proposals for three-photon gates}. (a-b) A qubit and qutrit \texttt{TOFFOLI} gate, each requiring four ancilla photons. (c) A qubit \texttt{FREDKIN} gate implemented with four ancilla photons.}
  \label{fig:tof_fred}
\end{figure*}
The PI teleportation scheme operates as follows. The qubit state of the incoming photon is first split according to its polarization. For each polarization component, a photon pair can be generated such that one of the pair’s photons is path-identified (i.e., overlapped) with the incoming photon. The incoming photon together with the photon in the overlapping path are then jointly detected, while the remaining photon passes through a polarizing beam splitter and is directed to the remote location (Fig.~\ref{fig:teleport} (c)). Since only one photon enters the gate, we know only one of the two detectors can fire. By conditioning on detection events at both detectors, we ensure that exactly one photon-pair creation process has occurred — the specific pair being determined by the incoming photon (Fig.~\ref{fig:teleport} (c)). With this logic, the outgoing photon carries the quantum information of the incoming photon, despite never being at the same location as the incoming photon.

This mechanism is closely related to the original work on path identity by Zou, Wang, and Mandel~\cite{zou1991induced, wang1991induced}. The experimental implementation of teleportation based on path-identity in Fig.\ref{fig:teleport}b does not allow to spatially separate the incoming and outgoing photons due to the requirement of overlapping the paths of the photons. However, the setup can be modified: instead of preventing the path information from ever being created (as in path-identity experiments), one can erase it at a later stage of the process. This modification introduces loss but enables genuine spatial separation in this new form of teleportation-like experiment, which we call \textit{path-information erasure teleportation} (PE teleportation). For higher-dimensional implementations, the scheme requires more ancillary photons.

Returning to the \texttt{SWAP} gate, we can now use the new forms of 
\textit{PI teleportation} or \textit{PE teleportation} twice (once from $A$ to $B$ and once from $B$ to $A$), see Fig.~\ref{fig:swap}. For higher dimensions, we also discover implementations that do not rely on two independent processes, but instead reduce the number of ancilla photons by exploiting additional coherent superpositions of photon-pair origins.

\section{\texttt{CNOT} gates}
Controlled-\texttt{NOT} operations are among the most fundamental multi-photon gates~\cite{barenco1995elementary}, where the state of one photon changes the state of the other. In the case of high-dimensional gates, it is useful to describe the gate as a controlled-\texttt{X} gate, where the \texttt{X} gate is a modulo-addition~\cite{babazadeh2017high,brandt2020high,meng2024experimental}, 
\begin{equation}
\widehat{\mathrm{CX}} \, |c, t\rangle
  = |c\rangle \hat{X}^c |t\rangle
  = |c\rangle |(t+c) \bmod d\rangle
\end{equation}
where the control $c$ and the target $t$ are qubit (or qudit) states, and $d$ is the dimension of the target qudit $t$.

Early demonstrations of photonic \texttt{CNOT} gates exploited shared Bell states between the control and target photons, as in the work of Gasparoni et al.~\cite{gasparoni2004realization}. Since then, numerous realizations of \texttt{CNOT} gates assisted by entangled~\cite{zhao2005experimental, huang2004experimental, gao2010teleportation, zeuner2018integrated} or single-photon ancillas~\cite{bao2007optical, li2021heralded} have been reported, often relying on Bell-state measurements. Moreover, several blueprints for generalized photonic control gates in higher dimensions and with multiple photons have been proposed, encoding the desired quantum transformations into complex entangled ancillary states~\cite{gao2020computer}.

Using \texttt{PyTheus}, we have discovered solutions for \texttt{CNOT} gates with control space $d_c=2$ and target space $d_t=2,3,4$, as well as a genuine 3-dimensional \texttt{CNOT} with $d_c=d_t=3$ (see Fig.~\ref{fig:cnot}). A previously introduced 3-dimensional \texttt{CNOT} required the generation of an intricate high-dimensionally entangled six-photon ancilla state~\cite{gao2020computer}, while our solution requires only four ancilla photons, generated from unentangled photon pairs.

The principle of the simplest, two-dimensional solution can be understood as follows. If the control photon is in state $\ket{0}$, it is directed to the ancilla detector, and an additional photon pair is generated between the second ancilla and the output. In this case, the target photon goes to its output and remains unchanged. If the control photon is in state $\ket{1}$, it is routed directly to the output; in that case, the target photon goes to the ancilla and another pair is generated with opposite states. These two options define the action of the \texttt{CNOT} in 2 dimensions. High-dimensional generalizations use the same principles. Based on this insight, we were able to conceptualize a generalization of high-dimensional \texttt{CNOT} gates of arbitrary dimension -- demonstrating that we have understood the underlying AI-discovered concept~\cite{de2005contextual, de2017understanding, krenn2022scientific}.

\section{\texttt{TOFFOLI} gate}
The controlled-controlled-\texttt{X} (\texttt{TOFFOLI}) gate is a fundamental three-photon gate: it advances the target only when both control photons are in the state $\ket{1}$ \cite{li2022chip, li2022quantum, dong2024experimental}. It is universal for reversible classical computation and forms the basis of arithmetic and oracle constructions in many quantum algorithms. Its operation is defined as
\begin{equation}
\begin{aligned}
\widehat{\mathrm{CCX}} \, |c_1,c_2,t\rangle
  &= |c_1,c_2\rangle\, \hat{X}^{(c_1 \land c_2)} |t\rangle \\
  &= |c_1,c_2\rangle\, \bigl|(t + (c_1 \land c_2)) \bmod d \bigr\rangle .
\end{aligned}
\end{equation}
Here, $c_1,c_2 \in {0,1}$ are the control qubits, $t$ is the target qubit (or qudit), $d$ is the dimension of the target system, and $\hat{X}$ denotes the generalized Pauli-$X$ (modulo-addition) operator. The symbol $\land$ represents the logical AND, which evaluates to $1$ if and only if both controls are $1$, and $0$ otherwise.

\texttt{PyTheus} discovers solutions for the \texttt{TOFFOLI} gate for $c_1,c_2\in\{0,1\}$ and $t \in\{0,1\}$ (which is the standard qubit case) and for $t \in\{0,1,2\}$, which is a generalized high-dimensional case. The solutions are displayed in Fig.~\ref{fig:tof_fred}(a) and (b). All solutions share the same properties: The three photons never meet at a single location, and there are no entangled photon pairs shared between the photons. Rather, the shared resource is the indistinguishability of the origins of the photon pair's origin. Surprisingly, despite these critical resource constraints, it is possible to achieve these goals with just four ancillas.

\section{\texttt{FREDKIN} gate}
The Fredkin gate (also known as controlled-\texttt{SWAP} or \texttt{CSWAP}) is another fundamental three-photon gate: it swaps the two target qubits only when the control is in state $\ket{1}$~\cite{smolin1996five, patel2016quantum, li2022quantum}. It is important for the SWAP test used in overlap and fidelity estimation~\cite{barenco1997stabilization, buhrman2001quantum} and for data-dependent routing of quantum information without measurement~\cite{weiss2024quantum, wang2025hardware}. Its operation is defined as
\begin{equation}
\begin{aligned}
\widehat{\mathrm{CSWAP}} \, \ket{c,a,b}
  &= \ket{c}\, \widehat{\mathrm{SWAP}}^{\,c} \ket{a,b} \\
  &=
  \begin{cases}
    \ket{0,a,b}, & c=0, \\[4pt] 
    \ket{1,b,a}, & c=1,
  \end{cases}
\end{aligned}
\end{equation}
Here $c,a,b\in\{0,1\}$ are qubit states (we restrict to $d=2$), and 
$\widehat{\mathrm{CSWAP}}$ is the controlled-SWAP operator, where 
$\widehat{\mathrm{SWAP}}$ (or $\hat{S}$) is defined by its action 
$\hat{S}\,\ket{x,y}=\ket{y,x}$ on computational-basis states.

Using \texttt{PyTheus}, we discover an experimental implementation for the 
\texttt{FREDKIN} gate that requires only four ancillas. The solution is shown in Fig.~\ref{fig:tof_fred}(c). As before, none of the photons meet at a single location (i.e., they do not interact) and no entangled pairs are shared among the photons.

\section{Outlook}
We have shown that the idea of indistinguishability by path identity is not only a resource for quantum interference and state generation, but also a powerful principle for the design of new forms of quantum gates. Accidentally, while conceptualizing the solutions, we discover a new analogue to quantum teleportation -- which is by itself an interesting finding. The presented experimental blueprints only use previously demonstrated technologies which indicates that they are also experimentally feasible. An interesting future question is how path identity and the solutions presented in this paper can be used for making quantum networks more efficient.

The unique way in which indistinguishability is generated via path identity -- not by erasing the information post-hoc, but by never letting it be generated -- might enable new experiments in causally indefinite processes, e.g. for superpositions of orders of quantum gates and~\cite{chiribella2013quantum, procopio2015experimental, rozema2024experimental, deng2025generalized}.

The work presented here is not only interesting for showing potentially new directions in quantum optics, but also raises interesting questions for AI-driven science: All solutions presented in this manuscript (except for the generalizations) were discovered by \texttt{PyTheus}, a system that can automatically discover new experiments in quantum optics. The generalizations of the AI-designed experiments were discovered by the human part of the research team. In principle, generalizations could also be discovered in an automated way -- for example, through the concept of meta-design~\cite{arlt2024meta} -- which would significantly accelerate research like that presented here. One important question still remains open: Could automated systems not only execute human-defined research, but also automatically generate the initial research questions and ideas themselves? 

\section*{Acknowledgments}
MK acknowledges support by the European Research Council (ERC) under the European Union’s Horizon Europe research and innovation programme (ERC-2024-STG, 101165179, ArtDisQ) and from the German Research Foundation DFG (EXC 2064/1, Project 390727645). X.G. acknowledges support from the NOA Collaborative Research Center and the Alexander von Humboldt Foundation.

\printbibliography
\end{document}